\documentclass[12pt]{article}

\usepackage{scicite}

\usepackage{epsfig}
\usepackage{times}

\def\prd{Phys.~Rev.~D}%
\newcommand{\fluxunits}{cm$^{-2}$\,s$^{-1}$}
\def\aap{{\em Astron. Astrophys.}}
\def\mnras{{\em Mon. Not. R. Astron. Soc.}}
\def\nat{{\em Nature}}
\def\apj{{\em Astrophys. J.}}
\def\apjs{{\em Astrophys. J. Suppl. Ser.}}
\def\apjl{{\em Astrophys. J.}}
\def\aj{{\em Astron. J.}}

\def\arcmin{\hbox{$^\prime$}}

\newenvironment{sciabstract}{%
\begin{quote} \bf}
{\end{quote}}

\newcounter{lastnote}

\title{Detection of Gamma Rays From a Starburst Galaxy} 

\author{
The H.E.S.S. Collaboration\footnote{To whom
correspondence should be addressed; E-mail: nedbal@ipnp.troja.mff.cuni.cz}
\\
\footnotesize{The full author list with affiliations can be found at the 
end of this paper}}

\date{}

\begin{document} 




\maketitle


\begin{sciabstract}
  \boldmath
  Starburst galaxies exhibit in their central regions a highly
  increased rate of supernovae, the remnants of which are thought to
  accelerate energetic cosmic rays up to energies of $\sim$
  10$^{15}$~eV. We report the detection of gamma rays -- tracers of
  such cosmic rays -- from the starburst galaxy NGC~253 using the
  H.E.S.S. array of imaging atmospheric Cherenkov telescopes. The
  gamma-ray flux above 220~GeV is $F= (5.5\pm 1.0_{\rm stat}\pm
  2.8_{\rm sys})\times 10^{-13}$ \fluxunits, implying a cosmic-ray
  density about three orders of magnitude larger than that in the
  center of the Milky Way. The fraction of cosmic-ray energy channeled
  into gamma rays in this starburst environment is 5 times larger than
  that in our Galaxy.  \unboldmath
\end{sciabstract}




%
%
Starburst galaxies are characterized by a boosted formation rate of
massive stars and an increased rate of supernovae in localised
regions, which also exhibit very high densities of gas and of
radiation fields. Their optical and infrared luminosity is dominated
by radiation from numerous young massive stars, most of which later
explode as supernovae. Given that most cosmic rays in normal galaxies
are expected to be accelerated in supernova remnants
\cite{Aharonian04}, starburst regions represent a favorable
environment for the acceleration of cosmic rays, resulting in orders
of magnitude higher cosmic ray energy densities compared to the local
value in our galaxy [e.g. \cite{Voelk89}]. Cosmic ray protons can
produce gamma radiation by inelastic collisions with ambient gas
particles and subsequent $\pi^0$-decay. Primary and secondary
cosmic-ray electrons can also produce gamma radiation by
Bremsstrahlung and up-scattering of low-energy photons from massive
stars or from ambient radiation fields. Starburst galaxies are
therefore considered promising sources of gamma-ray emission
\cite{Voelk96,Paglione96}. Here we report the detection of very high
energy (VHE; $>$ 100 GeV) gamma rays from the starburst galaxy
NGC~253.

%
%
NGC~253, at a distance of 2.6 to 3.9 Mpc
\cite{Puche88,Rekola05,Distance}, is one of the closest spiral
galaxies outside of the Local Group. It is similar to our Milky Way in
its overall star formation rate.  Its nucleus, however, is a starburst
region \cite{Engelbracht98} of very small spatial extent (few 100~pc),
characterised by a very high star formation rate per volume and thus
also by a very large mechanical energy production in form of supernova
explosions. Star formation activity is estimated to have been going on
for 20 -- 30 million years \cite{Engelbracht98} and can therefore be
considered to be in a steady state for the time scales governing
cosmic ray transport.
A supernova rate of $\sim$~0.1~yr$^{-1}$ has been inferred for the
entire galaxy from radio \cite{Antonucci88} and infrared
\cite{VanBuren94} observations.  The rate is most pronounced in the
central starburst region, where a conservative estimate yields a rate
of supernovae $\sim$~0.03~yr$^{-1}$, which is comparable to that in
our Galaxy \cite{Engelbracht98}. This suggests a very high local
cosmic ray energy density. The mean density of the interstellar gas in
the central starburst region is $n\simeq 600$~protons~cm$^{-3}$
\cite{Sorai00}, which is almost three orders of magnitude higher than
the average density of the gas in the Milky Way. The thermal continuum
emission of NGC~253 peaks in the far infrared (FIR) energy band at
$\sim$100~$\mu$m with a luminosity that is $\sim$~5 times the total
radiation from our own Galaxy \cite{Rice88}. This FIR emission
originates from interstellar dust, which reprocesses starlight from
the numerous young massive stars. The emission is highly concentrated
towards the small central starburst nucleus. Therefore, the density of
the radiation field in the starburst region is about a factor $10^5$
larger than the average value in the inner 100~pc around the Galactic
Center. The activity of NGC~253 has been shown to be of a pure
starburst nature and not due to an active supermassive black hole
\cite{Brunthaler09,Bauer07}.
Observations of radio \cite{Carilli92,Heesen09} and thermal X-ray
emission \cite{Dahlem98,Bauer08} show a hot diffuse halo, consistent
with the existence of a galactic wind extending out to $\sim$~9~kpc
from the galactic plane that transports matter and cosmic rays from
the nucleus to inter-galactic space and reaches asymptotically a bulk
speed of $\sim$~900~km/s \cite{Zirakashvili06}.

Given its proximity and its extraordinary properties, NGC~253 was
predicted to emit gamma rays at a detectable level
\cite{Paglione96}. Recent calculations give similar results
\cite{Torres05,Rephaeli09}. Previously, only upper limits have been
reported in the gamma-ray range, in the MeV-GeV range by EGRET
\cite{Blom99}, and in the TeV range by H.E.S.S. [based on 28 hours of
  observation \cite{Aharonian05}] and by CANGAROO III \cite{Itoh07}.
We report the result of continued observations of NGC~253 with the
H.E.S.S. telescope system with a much larger data sample.  (See the
Supporting Online Material (SOM) for a description of the experiment
and the detection technique.)

%
%

We obtained observations in 2005, 2007 and 2008. After rejecting those
data that did not have the required quality, we analyzed 119 hours of
live-time data . Even with this extremely long exposure, the measured
VHE gamma-ray flux of NGC 253 is at the limit of the
H.E.S.S. sensitivity. Thus advanced image analysis techniques were
required to extract a significant signal on top of the uniform
background of local cosmic rays impinging on the Earth's atmosphere --
only one out of $10^5$ recorded air showers represents a gamma-ray
from NGC 253. We used the ``Model Analysis'' technique \cite{Analysis}
(SOM), based on which we detected an excess of 247 events from the
direction of NGC~253 above 220~GeV, corresponding to a statistical
significance of 5.2 standard deviations (Fig. 1).  The signal is
steady and stable (a fit over the period of three years to a constant
has a chance probability of 47\%). The source position is
$\alpha_{J2000}=0^{\mathrm h}47^{\mathrm m}33^{\mathrm s}.6 \pm
30^{\mathrm s}$ , $\delta_{J2000}=-25^{\circ}18' 8'' \pm 27''$
consistent with the position of the optical center of NGC~253
($\alpha_{J2000}=0^{\mathrm h}47^{\mathrm m}33^{\mathrm s}.1$,
$\delta_{J2000}=-25^{\circ}17' 18''$). The distribution of excess
events is consistent with the point spread function of the
H.E.S.S. instrument, implying a source size of less than $4.2\arcmin$
(at a 1$\sigma$ confidence level, see Fig. 2 for a comparison of the
angular distribution of the gamma events with a point-like simulated
signal). The integral flux of the source above the threshold of
220~GeV is $F(>220\, {\rm GeV}) = (5.5\pm 1.0_{\rm stat}\pm 2.8_{\rm
  sys})\times 10^{-13}$ \fluxunits. This corresponds to 0.3\,\% of the
VHE gamma-ray flux from the Crab Nebula \cite{Aharonian06a}; given the
well-known uncertainties in the diffusion part of the particle
transport properties as well as the only approximate knowledge of the
starburst parameters, it is consistent with the original prediction
\cite{Paglione96} (Fig. 3).

%
%

As an external galaxy detected in gamma rays which, as a key property,
does not contain a significant active galactic nucleus, NGC~253 is a
member of a class of gamma-ray emitters external to the Milky Way and
the associated Large Magellanic Cloud (LMC). These gamma-ray emitters
apparently produce their own cosmic ray population. Except for the
starburst, NGC~253 is a normal galaxy.  So far only the LMC, a small
and close satellite of the Milky Way, was detected in gamma rays with
the EGRET instrument \cite{Sreekumar92}. In contrast there exists a
class of external galaxies detected in gamma-rays whose emission is --
according to present knowledge -- exclusively due to an active
galactic nucleus (AGN), driven by a supermassive Black Hole in their
center. Their physical characteristics are quite distinct from normal
galaxies and not the subject of the discussion here.

%
%

The detection of NGC 253 in VHE gamma rays implies a high energy
density of cosmic rays in this system.  One can calculate a
corresponding cosmic ray density directly from the
H.E.S.S. observations. Assuming a dominant hadronic origin of the
gamma-ray emission, the spatial density $N_\mathrm{p}(>E_\mathrm{p})$
of the gamma-ray generating protons in the starburst region with an
energy exceeding $E_\mathrm{p} \approx 220 /0.17$~$\mathrm{GeV}
\approx 1300$~GeV is about $4.9 \times 10^{-12}$~cm$^{-3}$ for the
measured gamma-ray flux above 220 GeV, independent of the distance to
NGC~253. This is about 2000 times larger than the corresponding
Galactic cosmic ray number density at the Solar System. And it is
about 1400 times higher than the density at the center of our Galaxy
\cite{Aharonian06b}.  Taking $E_\mathrm{p}N_\mathrm{p}(>E_\mathrm{p})$
as a rough measure of the energy density of cosmic rays above energy
$E_\mathrm{p}$ in NGC~253, $E_\mathrm{p}N_\mathrm{p}(>E_\mathrm{p})
\approx 6.4$~eV cm$^{-3}$ for $E_\mathrm{p}>1300$~GeV. This is larger
than the entire cosmic-ray energy density in the Galaxy near the Solar
System which is dominated by GeV-particles.
 
%
%

Gamma-ray production represents one channel for conversion and loss of
cosmic rays at TeV energies. The time between inelastic collisions of
hadronic cosmic rays and target protons and nuclei at
$E_\mathrm{p}\approx 1300$~GeV is of the order of $10^5\, {\rm yr}$
for a mean gas density of about 600 protons ${\rm cm}^{-3}$. These
collisional losses compete with two other processes in starbursts:
spatial losses of particles convected out of the considered region by
the wind, and diffusive losses (see the SOM for a
summary of the cosmic-ray transport characteristics in NGC~253).
Because of the very high gas density in the nucleus of NGC 253, the
ratio of hadronic gamma-ray production to energy loss by transport is
considerably higher than for a galaxy like ours. In the Milky Way, the
$\sim$~1300 GeV gamma-ray generating charged particles encounter about
0.6~g~cm$^{-2}$ of matter before they escape, extrapolating results
from \cite{Swordy90}. Their mean free path for inelastic nuclear
collisions is equivalent to about 56 g cm$^{-2}$.  Therefore, the
Galactic ratio of gamma-ray production probability to the escape
probability of 1300 GeV particles is about $10^{-2}$. If the
cosmic-ray energy production in the starburst region of NGC~253 is in
equilibrium with losses caused by nuclear collisions, then, for the
measured gas density and supernova rate -- together with an assumed
cosmic-ray production efficiency of $10^{50}$~erg per event and a
production spectrum $\propto E^{-2.1}$ \cite{Voelk96,Aharonian05} --
the expected integral gamma-ray flux above 220 GeV would be $\approx
10^{-11}$ \fluxunits. The observed flux is smaller than this
calorimetric flux by a factor $\approx 5 \times 10^{-2}$ -- again
independent of the distance. Therefore, the starburst region is only
mildly calorimetric. For a comparison see \cite{Aharonian05,
  Thompson07}. Nevertheless, the numbers imply that the conversion
efficiency of protons into gamma rays in the starburst region of
NGC~253 exceeds that in our Galaxy by almost an order of
magnitude. This comparatively high efficiency has another consequence:
assuming that the remaining structure of NGC~253 is approximately the
same as in our Galaxy, then the starburst nucleus is about 5 times
brighter in VHE gamma-rays than the associated galaxy, and the
starburst nucleus should outshine the rest of NGC~253. This is
consistent with the detection of a H.E.S.S. point source (Fig. 1).

%
%

Given these results one may ask whether they have a wider significance
regarding the nonthermal particle population in the Universe. A
starburst galaxy such as NGC~253 is a potential model for a phase of
galaxy formation as well as for two-body galaxy-galaxy interactions,
especially in the dense environment of large galaxy
clusters. High-energy gamma-ray emission as a result of these
processes should accompany the thermal IR emission of such luminous
Infrared Galaxies. The galactic winds present in these systems are
expected to massively populate intergalactic space with
nucleosynthesis products and cosmic rays.

\vspace*{0.5cm}

{\footnotesize {\bf Acknowledgements:}
The support of the Namibian authorities and of the University of Namibia
in facilitating the construction and operation of H.E.S.S. is gratefully
acknowledged, as is the support by the German Ministry for Education and
Research (BMBF), the Max Planck Society, the French Ministry for Research,
the CNRS-IN2P3 and the Astroparticle Interdisciplinary Programme of the
CNRS, the U.K. Science and Technology Facilities Council (STFC),
the IPNP of the Charles University, the Polish Ministry of Science and 
Higher Education, the South African Department of
Science and Technology and National Research Foundation, and by the
University of Namibia. We appreciate the excellent work of the technical
support staff in Berlin, Durham, Hamburg, Heidelberg, Palaiseau, Paris,
Saclay, and in Namibia in the construction and operation of the
equipment.
}

\epsfig{file=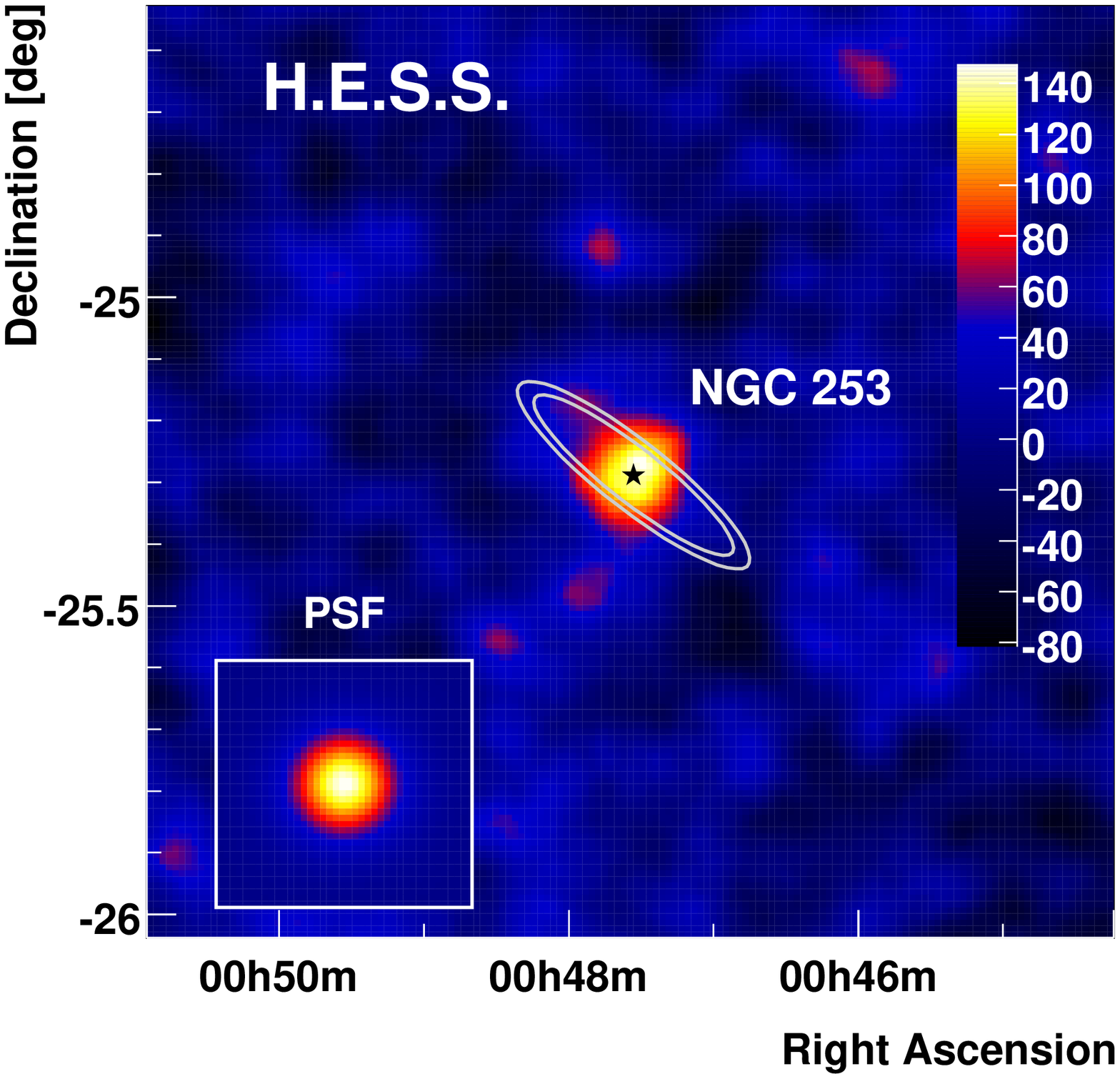,width=\textwidth}

\noindent {\bf Fig. 1.} A smoothed map of VHE gamma-ray excess of the
1.5$^\circ\times$1.5$^\circ$ region around NGC~253. A Gaussian with
RMS of $4.2\arcmin$ was used to smooth the map in order to reduce the
effect of fluctuations. The star shows the optical center of
NGC~253. The inlay represents an image of a Monte Carlo simulated
point source (i.e. the point-spread function of the instrument). The
white contours represent the optical emission of the whole NGC~253,
demonstrating that the VHE emission originates in the nucleus and not
in the disk. The contours correspond to constant surface brightness of
25 mag. arcsec.$^{-2}$ -- traditionally used to illustrate the extent
of the optical galaxy -- and 23.94 mag. arcsec.$^{-2}$ as given by
fits of \cite{Pence80}.

\epsfig{file=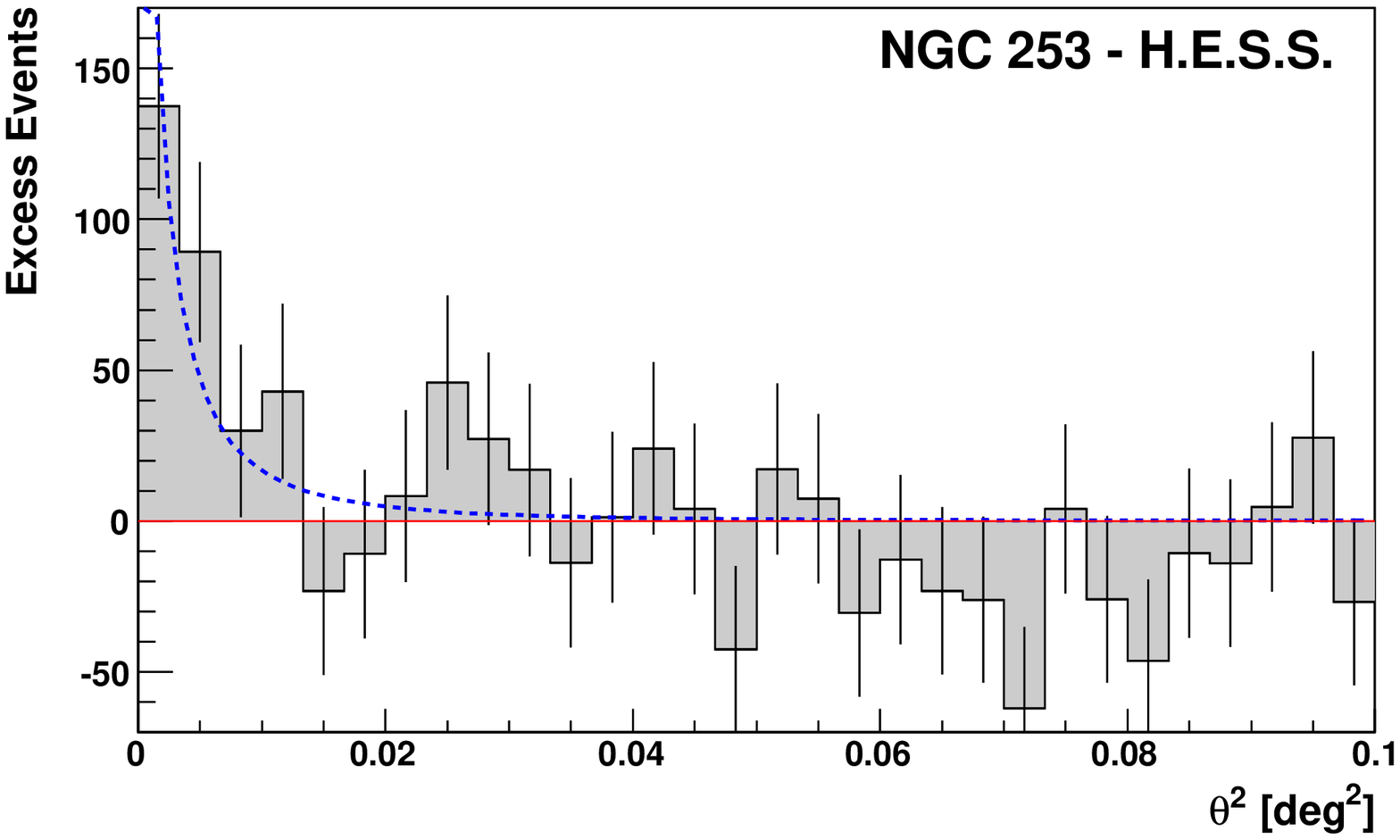,width=\textwidth}

\noindent {\bf Fig. 2.} Reconstructed directions of the gamma-ray like
events around NGC~253. $\theta$ denotes the angular distance between
the arrival direction and the position of the object. The background
estimated from off source regions, is uniform in the $\theta^2$
representation and has been subtracted here. The signal is consistent
with a H.E.S.S. point source (blue dashed line) corresponding to
$\theta < 4.2\arcmin$ or $< 3.2$~kpc at a distance of 2.6~Mpc.

\epsfig{file=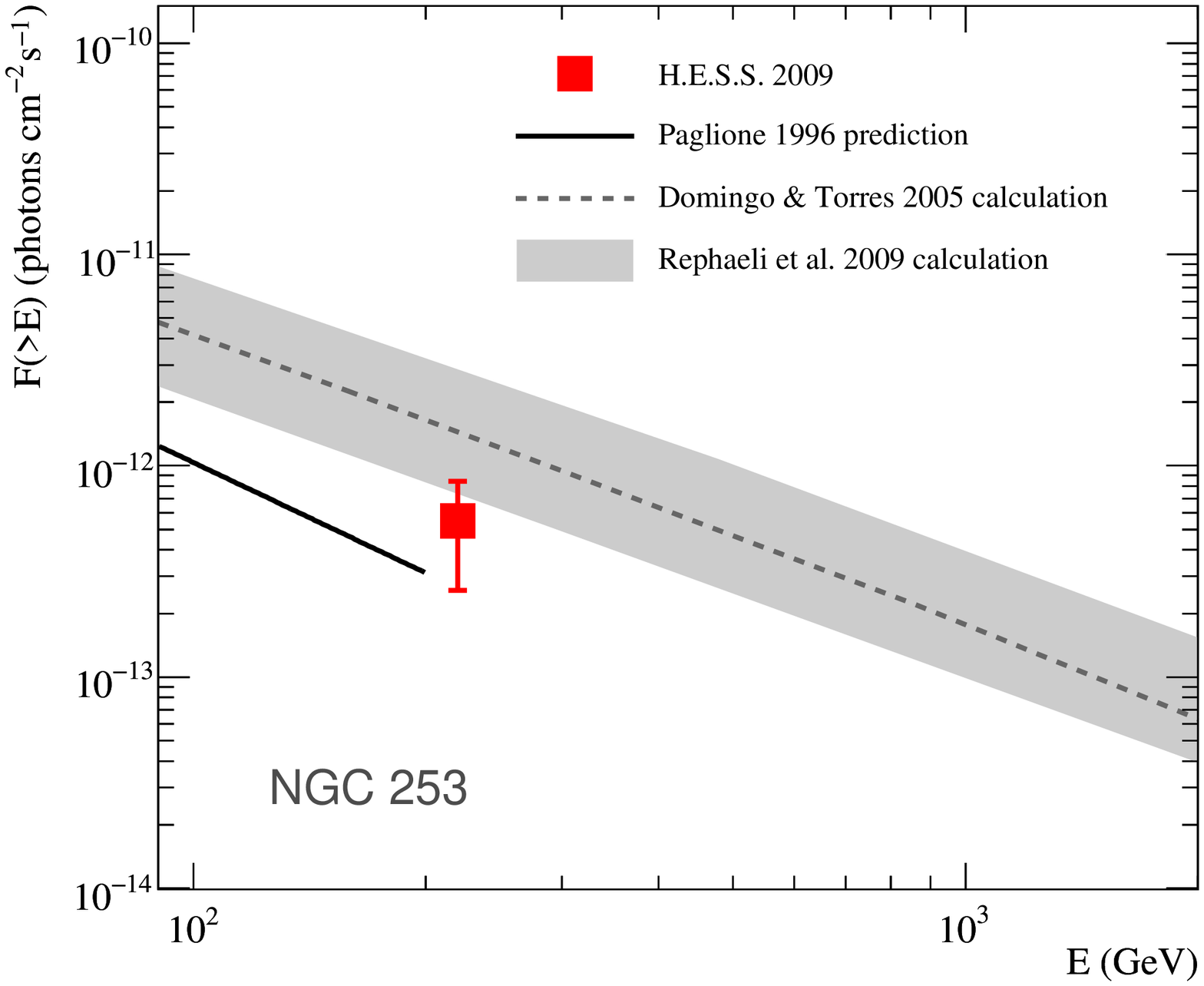,width=\textwidth}

\noindent {\bf Fig. 3.} The observed integral flux of gamma rays from
NGC~253 (red point) is compared to theoretical estimates
\cite{Paglione96,Aharonian05,Torres05}. The solid line corresponds to
the prediction by \cite{Paglione96}. The dashed line corresponds to
the model \cite{Torres05}. The grey-shaded band denotes the estimate
\cite{Rephaeli09}. The error of the H.E.S.S. measurement includes
systematic errors.

\clearpage

%

\bibliography{scibib}

\begin{thebibliography}{99}
\bibitem{Distance}
The following description is based on a distance of 2.6 Mpc, the final conclusions are,
  however, independent of distance.
\bibitem{Analysis} The ``Model Analysis'' is based on a comparison and
  fit of observed air shower images with a pre-computed library of
  images \cite{deNaurois06}; Another analysis technique, based on a
  machine learning algorithm called Boosted Decision Trees (BDT)
  \cite{Breiman84}, was used to verify the results. It was trained
  with simulated gamma-rays and with real cosmic-ray data from
  background fields \cite{Ohm09}. Both algorithms yield an improvement
  by a factor 1.5 to 1.7 in the statistical significance of faint
  sources compared to the standard image analysis \cite{Aharonian06a},
  as verified with a number of other gamma-ray sources. The ``Model
  Analysis'' result is found to be consistent with the one obtained
  with the ``BDT'' analysis. Details are given in the SOM.
\end{thebibliography}

\bibliographystyle{Science}


\subsection*{Supporting Online Material}
www.sciencemag.org \\
Materials and Methods \\
Supporting text \\
Figs.~S1, S2 \\
References \\


\newpage
\section*{Supporting Online Material}

\subsection*{Detection technique}
H.E.S.S. is an array of four imaging atmospheric Cherenkov telescopes
(IACTs) located in the Khomas highlands in Namibia [see \cite{sHinton04} for a
detailed description of the system]. IACTs observe the Cherenkov light
produced by relativistic particles in electromagnetic showers resulting from
gamma-ray collisions in the atmosphere, thus using the Earth's atmosphere as
a detector medium. The optical Cherenkov light is detected with an ultra-fast
optical camera assembled from photomultiplier pixels that takes images of the
showers. Gamma rays are separated from the charged cosmic rays on the basis
of different shower properties [e.g. \cite{sVoelk09}]. H.E.S.S. observes the
sky in the VHE gamma-ray regime. With the standard analysis it can detect a
source of 1\% of the flux of the ``standard candle'' Crab Nebula in 25 hours
at a 5$\sigma$ level \cite{sAharonian06}.
  
\subsection*{Data set and data quality}
The large dataset together with the weak signal of NGC~253 require careful
control of the quality of the data used for analysis. In order to reduce
the effect of hardware problems and atmospheric fluctuations on the flux
determination, a system of quality criteria cuts is applied to reject the data
with bad or dubious quality [see also discussion in \cite{sAharonian06}].

A substantial part of the systematic error of the flux is caused by atmospheric
conditions. Cloudy sky and dust in the atmosphere absorb the Cherenkov light
from the electromagnetic shower, which results in a lower detection rate and in
an underestimation of the flux. The rate at which cosmic-ray showers trigger
the instruments provides a sensitive monitor for atmospheric transparency. As a
function of the zenith angle of telescope pointing, a nominal trigger rate is
estimated for each 28-minute observation run. If the real trigger rate is less
than 80\% of the predicted value, the run is rejected. Additionally, to avoid
intra-run fluctuations caused by e.g. clouds, a cut on the RMS of the trigger
rate within one run is applied (max. 10\% is allowed). Further cuts detailed by
\cite{sAharonian06} are applied in order to  limit the effect of malfunctioning
pixels in the camera and the correct pointing and tracking of the telescopes as
well as to assure a proper calibration of each run.

NGC~253 was observed with the full array of four H.E.S.S. telescopes during
2005, 2007 and 2008 for a total of 192 hours. After applying the quality
selection cuts and correcting for the instrument dead-time, 119 hours of
good-quality live-time data is used for the analysis. The average zenith angle
of these observations was 11$^\circ$ resulting in an energy threshold of
220~GeV for the ``Model analysis'' and 260~GeV for the BDT analysis, after all
cuts applied in the respective analyses.

\subsection*{``Model" image analysis}
The main results presented here were produced with the {\it Model Analysis}
\cite{sdeNaurois09}. Conventional analysis techniques rely on an image cleaning
procedure to extract relevant information in the camera, and reduce the shower
information to a few parameters through a parametrization of the image
shape. In contrast, the Model Analysis is based on a pixel-per-pixel comparison
of the actual recorded intensity of a signal in the camera with a pre-calculated
shower model, without any image cleaning or parametrization. The shower model
is based on a semi-analytical description of shower development in the
atmosphere. Template images for a large range in primary energy, zenith angle,
impact distance and depth of first interaction are generated using a dedicated
code, stored into files, and used in the comparison with the actual images. The
noise distributions in the pixel due to the night sky background is taken into
account in the model fit. A precise description of statistical fluctuations is
used in form of a log-likelihood minimization, which results in a superior
treatment of shower tails. Therefore, the Model Analysis results in a more
precise reconstruction and a better background suppression than more
conventional techniques, thus leading to an improved sensitivity. The remaining
background in either method is estimated using reflected control regions \cite{sAharonian06}.

A goodness-of-fit approach is chosen to compare the model prediction
and the actual shower images, in order to check the compatibility of
the recorded events with a pure gamma-ray hypothesis and therefore
provide gamma-hadron separation.  The goodness-of-fit is defined as a
normalised sum over all pixels of the difference between the actual
pixel log-likelihood (at the end of minimization) and its expectation
value, taking into account Poisson fluctuations of the number of
photons, electronic and night sky background fluctuations. The
goodness-of-fit parameter retains 70\% of gamma-rays and rejects more
than 95\% of background events, yielding a quality factor 
\begin{equation}
Q =\frac{\epsilon_\gamma }{\sqrt {\epsilon_\mathrm{hadrons}}} \approx 4
\end{equation}

Additional discriminating parameters, such as the depth of first
interaction, are also used in the Model Analysis to improve the
gamma-hadron separation.  Cross-checks performed on several sources
(Crab Nebula, PKS 2155-305, and others) show that, on average, the
Model Analysis yields a gamma-ray efficiency in the order of 2.5 times
higher than that of the Hillas reconstruction method with hard cuts
(used in most H.E.S.S. publications), with a similar increase in the
background level (depending in particular on the source spectral
index). The corresponding improvement of sensitivity is by a factor
of slightly less than 2.

\subsection*{Analysis using Boosted Decision Trees}
The results  were cross-checked  using an analysis  based on an
independent calibration procedure and a different background discrimination
method. It is based on a machine learning algorithm called Boosted Decision
Trees (BDT) \cite{sBreiman84,sOhm09}. This mathematical tool was also utilized for particle
identification in high-energy physics [e.g. \cite{sYang05,sAbazov08}]
and was applied recently in astrophysics for e.g. supernova
searches \cite{sBailey07}.  A similar approach called Random Forest
\cite{sBreiman2001}, which is also based on decision trees, was utilized in
ground-based VHE astronomy \cite{sBock2004,sAlbert2008a,sEgberts2008}. 
The BDT method uses various parameters as input for the
classification. Four parameters are of the classical Hillas-type
\cite{sHillas85}, where the measured width and length of a shower image
is compared to the expectation for Monte-Carlo $\gamma$-rays and hadronic
cosmic rays (MRSW, MRSL, MRSWO, MRSLO \cite{sOhm09}). Additionally, the
averaged spread in reconstructed energy between triggered telescopes
$\Delta$E~/~E and the depth of the shower maximum X$_{\mathrm{max}}$ are
used as input parameters \cite{sOhm09}.
It returns a
continuous variable classifying the event, with gamma-ray like events at one
end of the range and cosmic-ray background like events at the other end. Monte
Carlo simulations of gamma rays and hadronic cosmic rays from observations
without any gamma-ray emitter in the field-of-view were used to teach the BDT
the differences between the two.
After the training of the BDT, it was used to classify
events of the NGC~253 observations. Thus, the hadronic background
in the data set could be reduced by a factor of 2 compared to the H.E.S.S. 
standard analysis \cite{sAharonian06}, giving a 1.5 times higher significance
and facilitating the detection of NGC~253 in VHE gamma-rays.

Using the BDT method, an excess of 283 (compared to 247 for the Model analysis)
gamma rays above the threshold of 260~GeV is obtained, corresponding to a
statistical significance of 4.9 $\sigma$ (compared to 5.2 $\sigma$ for the
Model analysis). The flux value derived using this method  is $F(>260\, {\rm
  GeV}) = (5.7\pm 1.3_{\rm stat} \pm 2.8_{\rm sys})\times
10^{-13}$\fluxunits\ and confirms the value of the ``Model analysis''. In order to perform the
comparison, one needs to extrapolate the BDT value to 220~GeV -- the threshold
of the ``Model analysis.'' The extrapolated values are $6.8\times 10^{-13}$ \fluxunits\ and $7.3\times 10^{-13}$ \fluxunits, respectively, assuming a power law with a spectral index
of 2.1 and 2.5, respectively. These two indices are considered to encompass a realistic range of values. Both extrapolated flux values are well within the errors of the ``Model analysis''
result ($(5.5\pm 1.0_{\rm stat}\pm 2.8_{\rm sys})\times 10^{-13}$
\fluxunits). The excess map is shown in Fig. S1 and the distribution of arrival
directions of gamma rays around NGC~253 is depicted in Fig. S2. The
agreement between Monte Carlo simulations and real data is shown for
the methods in Fig. S4 and Fig. S5.

\subsection*{H.E.S.S. standard analysis}
The H.E.S.S. standard analysis is based on a simple parametrization of the
shower images by the Hillas parameters -- width and length. These parameters
are scaled to their nominal values for gamma rays and averaged over all the
images of a given shower. The scaled parameters are subsequently used for event
selection by means of cuts on these parameters and for reconstruction of
physical properties of the shower \cite{sAharonian06}.  However, the  technique does not fully exploit the information obtained by the finely-pixellated
cameras of modern IACT telescopes. Techniques which use the full pixel
information -- such as the Model analysis -- and which include additional
parameters and better account for correlations between the parameters -- such
as the BDT analysis -- provide an improved background rejection and are more
appropriate for analyzing the weakest sources. Using the standard analysis with
standard background rejection cuts and the reflected background estimation
\cite{sAharonian06}, an excess of 252 gamma rays is found, resulting in a
statistical significance of 3.2$\sigma$. This is in agreement with the
expectations for a flux level of 0.3\% of the Crab Nebula.

\subsection*{Systematic uncertainties} The flux determination is
influenced by several sources of systematic uncertainties. The most important
ones are uncertainties due to the atmospheric model used for Monte Carlo
simulation ($\sim$20\%), run-by-run variability and calibration ($\sim$20\%),
malfunctioning camera pixels ($\sim$10\%), background estimation ($\sim$15\%)
and selection cuts ($\sim$15\%). 
Additionally, the integral flux above a
threshold can be biased by the energy-scale uncertainty of $\sim$15\%. The
total error of the flux above a threshold is thus very conservatively estimated
as 40-50\%.

The only systematic factor that can affect the detection significance is the
uncertainty in background estimation, caused by a possible non-uniformity of
the background over the field-of-view (FOV). In order to investigate this
effect, a distribution of significances in off-source bins in the FOV is shown
in Fig. S3. In case of a uniform background, this distribution will be
Gaussian-distributed with mean of 0 and RMS of 1. The distribution in case of
NGC~253 is well fit by a Gaussian of mean 0.1 and RMS of 1.1 (Fig. S3). The
systematic error of the detection significance is constrained to approximately
10\%. Note that using an alternative background estimation with ring-shaped
control regions \cite{sAharonian06} results in a higher significance of
5.9~$\sigma$ (Model analysis) and 5.3~$\sigma$ (BDT analysis). The more
conservative values were thus used in deriving the results.

\subsection*{Cosmic ray transport in NGC~253}
{ 
Three distinct processes can lead to losses of
cosmic-rays in the nucleus of NGC~253. Inelastic collisions of hadronic cosmic-rays with protons
and nuclei of the thermal gas lead to pion production. In addition to that cosmic-rays
can also leave the starburst region convectively and diffusively. The timescale of the
convective transport of cosmic rays out of the starburst region
is $H/v \sim 10^5$~yr, where, for a distance of 2.6 Mpc, $H
\sim 60$~pc is the height of the roughly cylindrical starburst region
whose diameter is about 300~pc \cite{sWeaver02}, and $v \sim 500$~km/s
is the assumed mean wind speed. The time between inelastic
collisions of the hadronic cosmic rays and the protons and nuclei of the thermal gas at
$E_\mathrm{p}\approx 1300$~GeV is also of the order of $10^5\, {\rm
  yr}$ for a mean gas density of about 600 protons ${\rm cm}^{-3}$. In addition the particles can also
diffuse through the gas, being scattered by irregularities of the
magnetic field. This latter process depends on the particle energy,
with large uncertainties, but certainly increases with particle energy
\cite{sBlandford80, sAharonian05}.

The detailed description of the interaction of cosmic rays with the
gas is somewhat more complex than outlined above: the hot and rarefied
gas produced by individual supernovae creates a contiguous hot gas
component which pushes its way out from the starburst region between
the massive clouds of cold gas from which the stars form.  A high
temperature ($10^6 - 10^7$~K) outflow is observed well beyond the
starburst region \cite{sBauer08}. Cosmic rays are accelerated in shock
waves that also produce the hot gas. Low energy particles are largely
confined in this rarefied gas which tends to carry them out of the
starburst region. Such a confinement is not likely for the high energy
particles relevant for high-energy gamma-ray production. Hence they will
also penetrate the dense cold gas surrounding the hot gas flow. For
low energy particles this penetration is an open question.
 Cosmic-ray electrons  will in any case lose a major fraction of
their energy by Bremsstrahlung, IC collisions in the strongly enhanced radiation field
and by synchrotron radiation in the magnetic field of the starburst
region.
\clearpage

\epsfig{file=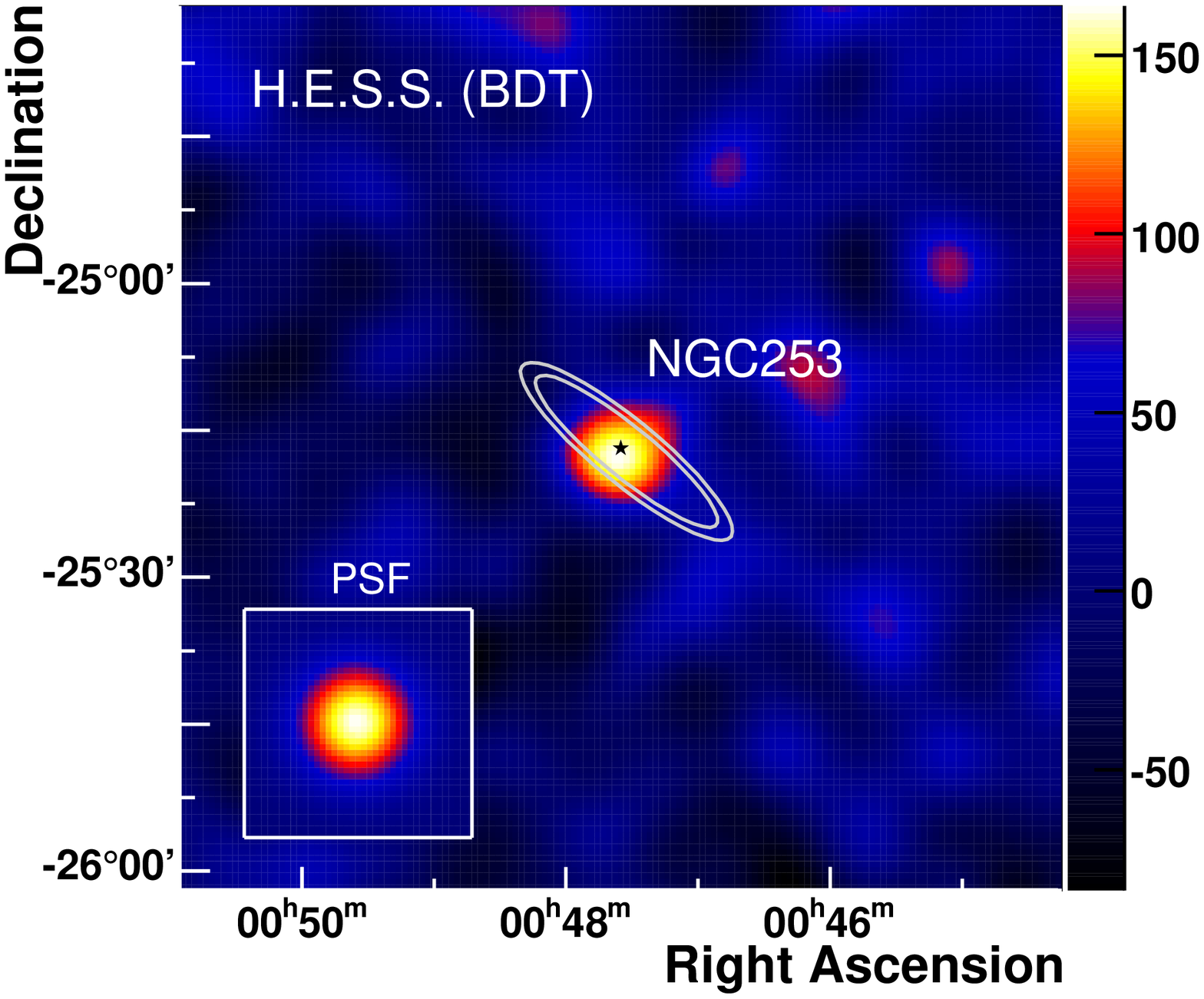,width=\textwidth}
\noindent {\bf Fig. S1.} A smoothed map of VHE gamma-ray excess of the
1.5$^\circ\times$1.5$^\circ$ region around NGC~253 obtained using the BDT method. A
Gaussian with RMS of $4.2\arcmin$ is used to smooth the map in order to reduce
the effect of fluctuations. 
The inlay represents an image of a Monte Carlo simulated point source
(i.e. the point spread function of the instrument). The white
  contours correspond to constant optical surface brightness
of 25 mag. arcsec.$^{-2}$ and 23.94 mag. arcsec.$^{-2}$ as given by fits
of \cite{sPence80}.

\epsfig{file=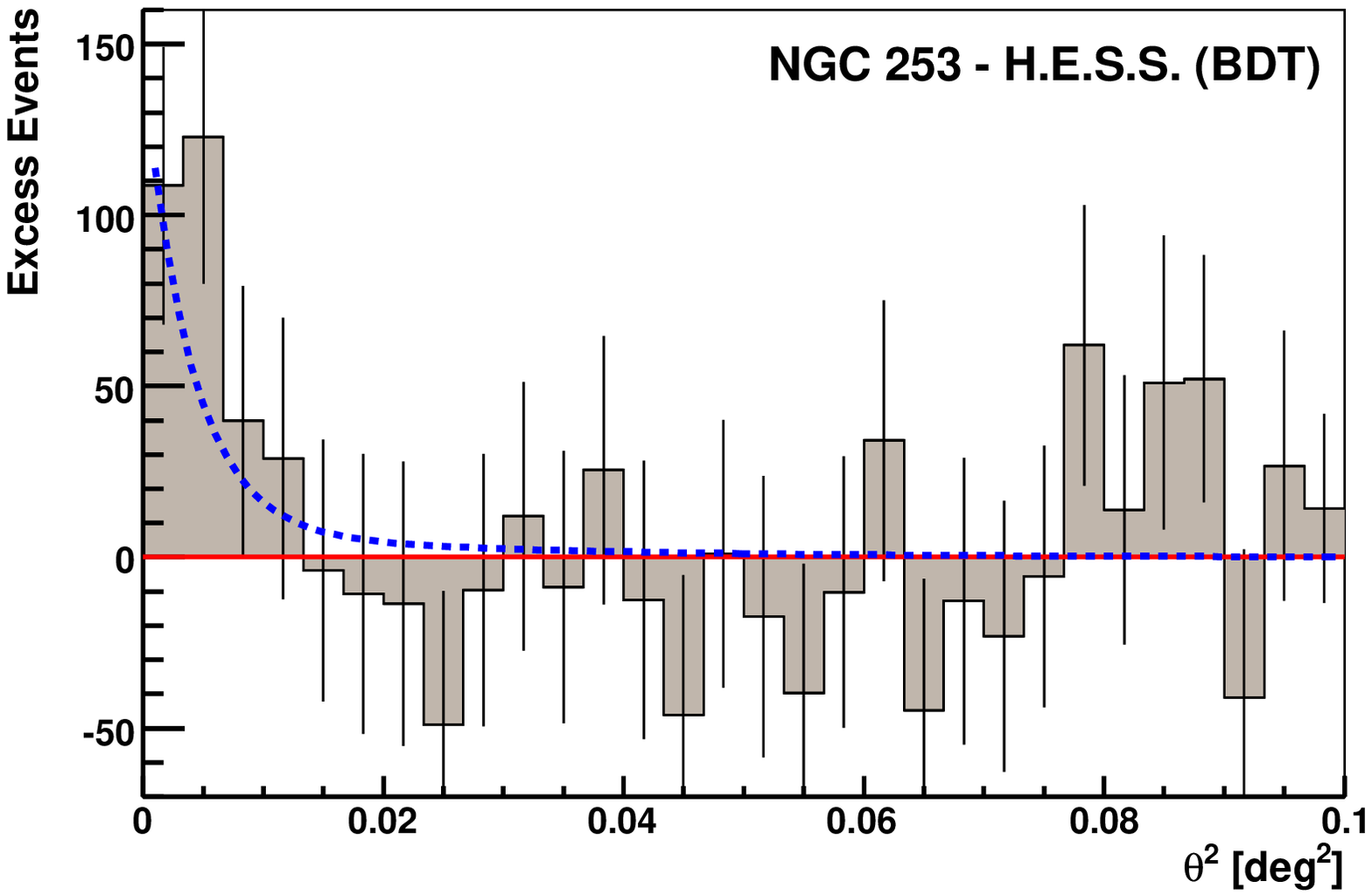,width=\textwidth}

\noindent {\bf Fig. S2.} Reconstructed directions of the gamma-ray
like events around NGC~253 using the BDT method. $\theta$ denotes the
angular distance between the arrival direction and the position of the
object. The signal is consistent with a H.E.S.S. point source (the dashed 
blue line shows how H.E.S.S. sees a Monte-Carlo simulated point source using 
the BDT method). The background was estimated from off source regions, 
is uniform in the $\theta^2$ representation and has been subtracted here.

\epsfig{file=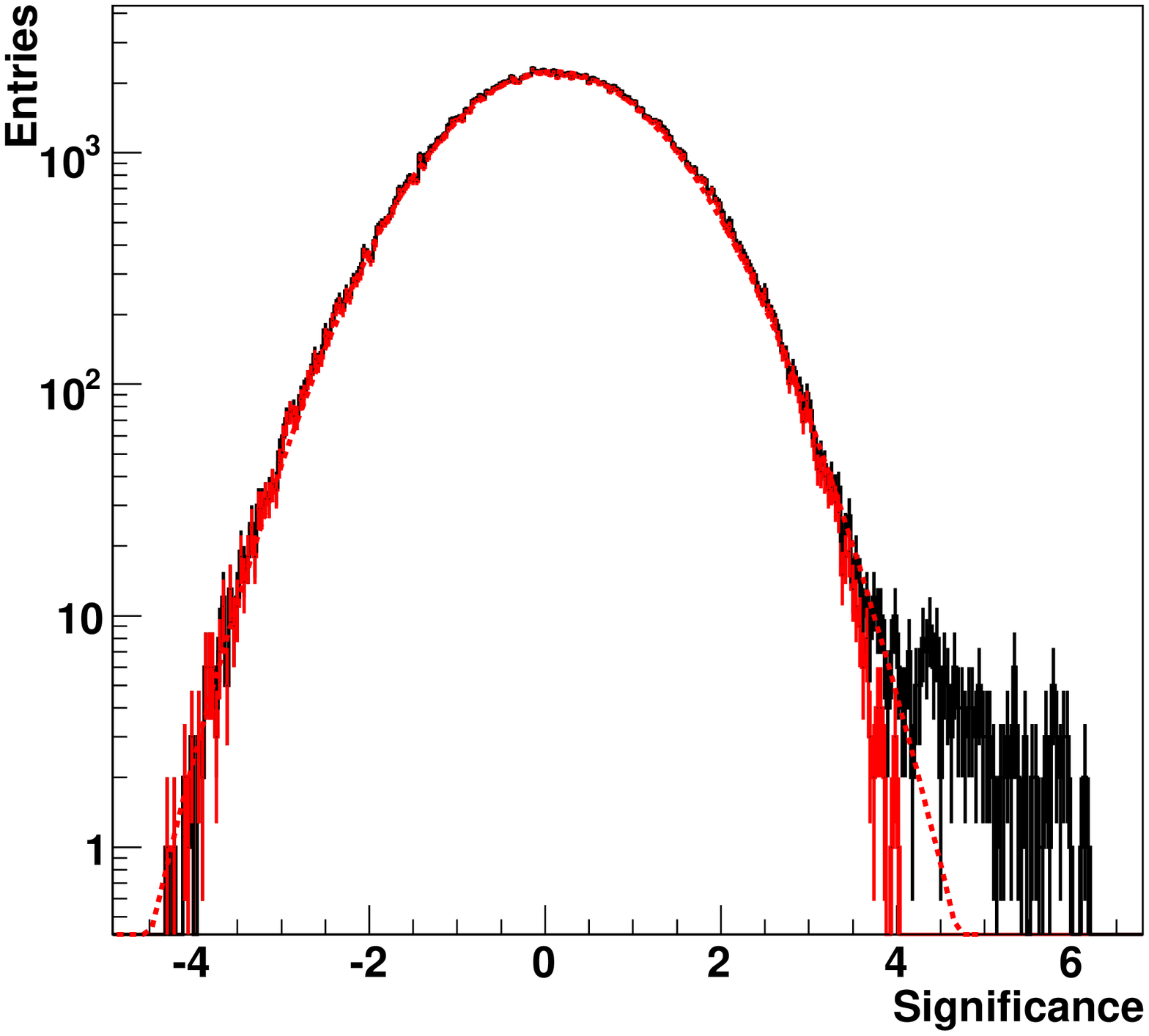,width=\textwidth}

\noindent {\bf Fig. S3.} Distribution of bin significances of the
  field-of-view including and excluding NGC~253 (black and red line,
  respectively) for the ``Model analysis.'' The distribution excluding
  the source is fitted by a Gaussian with a mean of 0.1 and an RMS of
  1.1 (red dashed line).

\epsfig{file=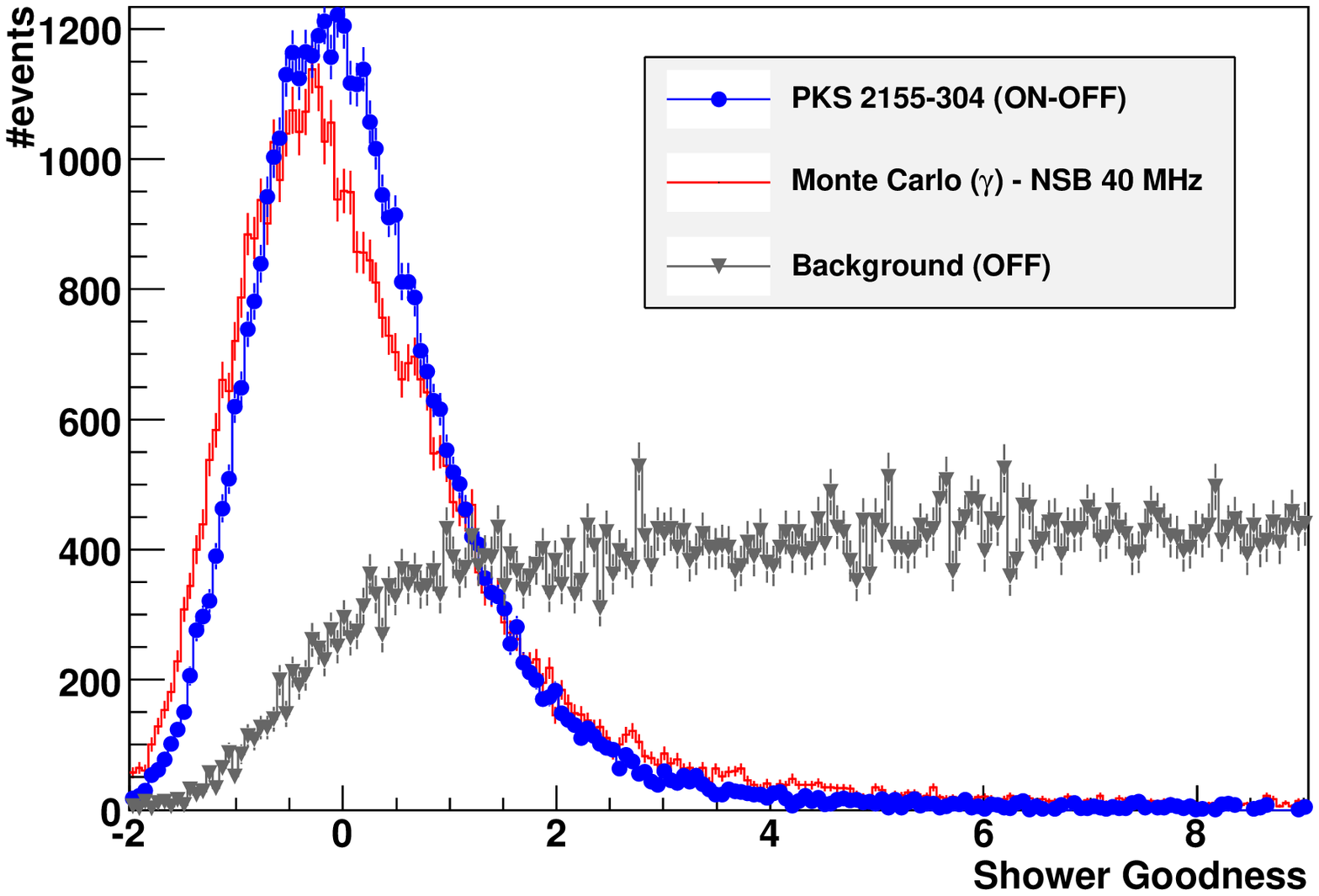,width=\textwidth}

\noindent {\bf Fig. S4.} Distribution of Shower Goodness for observational data 
taken on the blazar PKS~2155--304 (blue points indicate gamma-ray excess events, 
grey triangles cosmic-ray background events), compared with a simulation (red histogram) 
with a similar night sky background level. Figure taken from \cite{sdeNaurois09}.

\epsfig{file=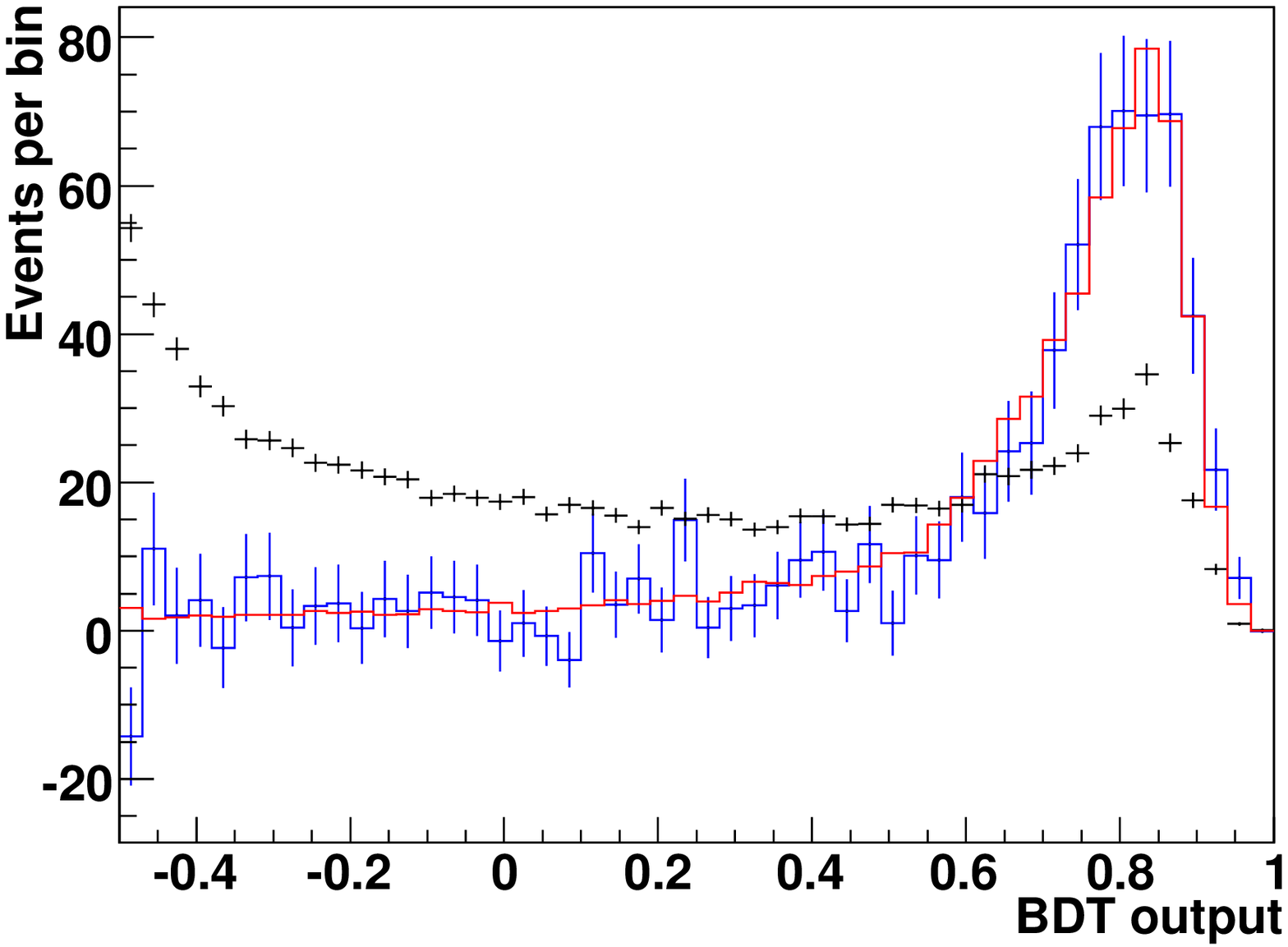,width=\textwidth}

\noindent {\bf Fig. S5.} Distribution of BDT output for observational data 
taken on the H.E.S.S. source HESS~J1745--290 (blue points indicate gamma-ray excess events, 
grey points cosmic-ray background events), compared with a simulation (red histogram). 
Figure adapted from \cite{sOhm09}.

\section*{Full list of authors}
\par\noindent{\large
F.~Acero$^{15}$,
F.~Aharonian$^{1,13}$,
A.G.~Akhperjanian$^{2}$,
G.~Anton$^{16}$,
U.~Barres de Almeida$^{8\dag}$,
A.R.~Bazer-Bachi$^{3}$,
Y.~Becherini$^{12}$,
B.~Behera$^{14}$,
K.~Bernl\"ohr$^{1,5}$,
A.~Bochow$^{1}$
% ...
}
\\
\\
{\footnotesize
\begin{enumerate}

\item Max-Planck-Institut f\"ur Kernphysik, P.O. Box 103980, D 69029
Heidelberg, Germany

\item Yerevan Physics Institute, 2 Alikhanian Brothers St., 375036 Yerevan,
Armenia

\item Centre d'Etude Spatiale des Rayonnements, CNRS/UPS, 9 av. du Colonel
Roche, BP
4346, F-31029 Toulouse Cedex 4, France

\item Universit\"at Hamburg, Institut f\"ur Experimentalphysik, Luruper
Chaussee
149, D 22761 Hamburg, Germany

\item Institut f\"ur Physik, Humboldt-Universit\"at zu Berlin, Newtonstr. 15,
D 12489 Berlin, Germany

\item LUTH, Observatoire de Paris, CNRS, Universit\'e Paris Diderot, 5
Place Jules Janssen, 92190 Meudon,
France

\item IRFU/DSM/CEA, CE Saclay, F-91191
Gif-sur-Yvette, Cedex, France

\item University of Durham, Department of Physics, South Road, Durham DH1
3LE,
U.K.

\item Unit for Space Physics, North-West University, Potchefstroom 2520,
    South Africa

\item Laboratoire Leprince-Ringuet, Ecole Polytechnique, CNRS/IN2P3,
 F-91128 Palaiseau, France

\item Laboratoire d'Annecy-le-Vieux de Physique des Particules,
Universit\'{e} de Savoie, CNRS/IN2P3, F-74941 Annecy-le-Vieux,
France

\item Astroparticule et Cosmologie (APC), CNRS, Universite Paris 7 Denis
Diderot,
10, rue Alice Domon et Leonie Duquet, F-75205 Paris Cedex 13, France
$^\ddag$

\item Dublin Institute for Advanced Studies, 5 Merrion Square, Dublin 2,
Ireland

\item Landessternwarte, Universit\"at Heidelberg, K\"onigstuhl, D 69117
Heidelberg, Germany

\item Laboratoire de Physique Th\'eorique et Astroparticules,
Universit\'e Montpellier 2, CNRS/IN2P3, CC 70, Place Eug\`ene Bataillon,
F-34095
Montpellier Cedex 5, France

\item Universit\"at Erlangen-N\"urnberg, Physikalisches Institut,
Erwin-Rommel-Str. 1,
D 91058 Erlangen, Germany

\item Laboratoire d'Astrophysique de Grenoble, INSU/CNRS, Universit\'e
Joseph Fourier, BP
53, F-38041 Grenoble Cedex 9, France

\item Institut f\"ur Astronomie und Astrophysik, Universit\"at T\"ubingen,
Sand 1, D 72076 T\"ubingen, Germany

\item LPNHE, Universit\'e Pierre et Marie Curie Paris 6, Universit\'e
Denis Diderot
Paris 7, CNRS/IN2P3, 4 Place Jussieu, F-75252, Paris Cedex 5, France

\item Charles University, Faculty of Mathematics and Physics, Institute of
Particle and Nuclear Physics, V Hole\v{s}ovi\v{c}k\'{a}ch 2, 180 00

\item Institut f\"ur Theoretische Physik, Lehrstuhl IV: Weltraum und
Astrophysik,
    Ruhr-Universit\"at Bochum, D 44780 Bochum, Germany

\item University of Namibia, Private Bag 13301, Windhoek, Namibia

\item Obserwatorium Astronomiczne, Uniwersytet Jagiello{\'n}ski, ul. Orla
171,
30-244 Krak{\'o}w, Poland

\item Nicolaus Copernicus Astronomical Center, ul. Bartycka 18, 00-716
Warsaw,
Poland

\item School of Physics \& Astronomy, University of Leeds, Leeds LS2 9JT, UK

\item School of Chemistry \& Physics,
 University of Adelaide, Adelaide 5005, Australia

\item Toru{\'n} Centre for Astronomy, Nicolaus Copernicus University, ul.
Gagarina 11, 87-100 Toru{\'n}, Poland

\item Instytut Fizyki J\c{a}drowej PAN, ul. Radzikowskiego 152, 31-342
Krak{\'o}w,
Poland

\item Astronomical Observatory, The University of Warsaw, Al. Ujazdowskie
4, 00-478 Warsaw, Poland

\item European Associated Laboratory for Gamma-Ray Astronomy, jointly
supported by CNRS and MPG

\item Institut für Astro und Teilchenphysik, Leopold-Franzens-Universität
Innsbruck, A6020 Innsbruck, Austria

\item KIPAC, Stanford University, Stanford, CA 94305, USA
\end{enumerate}

\begin{itemize}

\item[$\dag$]{supported by CAPES Foundation, Ministry of Education of Brazil}

\item[$\ddag$]{UMR 7164 (CNRS, Universit\'e Paris VII, CEA, Observatoire
de Paris)}

\end{itemize}
}

\end{document}